\documentclass[dvips]{article}

\usepackage{icrc2011}

\title{VHE Observations of Galactic binary systems with VERITAS}

\newcommand{\etal}{\MakeLowercase{\textit{et al. }}} 
\shorttitle{G.Maier  \etal Binary observations with VERITAS}

\authors{Gernot Maier$^{1}$ for the VERITAS Collaboration}
\afiliations{$^1$DESY, Platanenallee 6, 15738 Zeuthen, Germany\\ }
\email{gernot.maier@desy.de}

\abstract{
A few Galactic objects are known to be variable sources of photons with energies above 100 GeV. These systems are mostly binaries, where variability can often be connected to the orbital period. Particle acceleration and gamma-ray production processes in binaries are generally not well understood. We present here an overview of VERITAS observations of binary systems at very-high energies (VHE) with emphasis on LS I +61 303 and 1A0535+262. The results are discussed in the context of recent observations of RXTE, Swift and Fermi LAT at X- and Gamma-ray energies. 
}

\begin{document}
\maketitle

\section{Introduction}

Several high-mass X-ray binaries have been detected in the VHE energy range in the past years. 
These binaries constitute the only known class of galactic objects with variable point-like VHE emission.
The class currently contains three members: PSR B1959-63/LS 2883 \cite{Aharonian-2005b}, LS 5039 \cite{Aharonian-2005c}
 and LS I +61 303 \cite{Albert-2006, Acciari-2008}.
A fourth TeV binary might be the unidentified VHE source HESS J0632+057 \cite{Aharonian-2007},  a variable gamma-ray source located in the Monoceros region. 

The variable emission from VHE binaries is likely connected to changes in physical parameters associated with the orbital movement. 
This makes each of these systems unique, since orientation, sizes, and eccentricity of the orbits vary widely.
Orbital periods range from a a few days in LS 5039 to several years in PSR B1959-63. 
The changes during an orbit are not of geometrical nature only, beside distances and orientation of the objects towards each
other and the observer are variations in photon and matter densities around the acceleration sites 
the main causes for the variable emission.

PSR 1259-63 is currently the only VHE binary with a clearly identified compact object - a 48 ms pulsar orbiting a Be star in an eccentric orbit.
No pulsed emission at any wavelength has been detected from the binaries LS 5039 and LS I + 61 303, although  pulsed
signals might get absorbed in the dense photon field of the nearby bright companion stars (see e.g, \cite{Rea-2010}). 
In general, the VHE emission in these objects is explained by the acceleration of charged particles in shocks 
created by the collision of the wind of the stellar companion with the expanding pulsar wind or by acceleration 
in accretion-powered jets. The high-energy particles produce VHE photons in hadronic and/or leptonic 
(inverse Compton scattering of low-energy stellar photons) interactions.

We present in the following observations of two binary systems with VERITAS at energies above 100 GeV: 
 LS I +61 303 was observed for more than 60 hours during several orbital phases between 2008 and 2010
and 
the  Be/X-ray binary 1A0535+262 was observed during a giant X-ray outburst in December 2009 for a full orbit.
For details on the X-ray analysis and results, see \cite{Acciari-2011a, Acciari-2011b}.

\section{VERITAS}

VERITAS is an array of four imaging atmospheric-Cherenkov telescopes (IACT)
located at the Fred Lawrence Whipple Observatory
in southern Arizona.
It combines a large effective
area ( $>10^5$ m$^2$) over a wide energy range (100 GeV
to 30 TeV) with good energy (15-20\%) and angular
($\approx0.1^{\mathrm{o}}$) resolution.
The field of view of the VERITAS telescopes is 3.5$^{\mathrm{o}}$.
The high sensitivity of VERITAS enables
the detection of sources with a flux of 1\% of the Crab
Nebula in about 25 hours of observations.
For more details on the VERITAS
instrument, see e.g.~\cite{Holder-2011}.

\section{LS I +61 303}

\begin{figure}[!t]
  \vspace{5mm}
  \centering
  \includegraphics[width=0.9\linewidth]{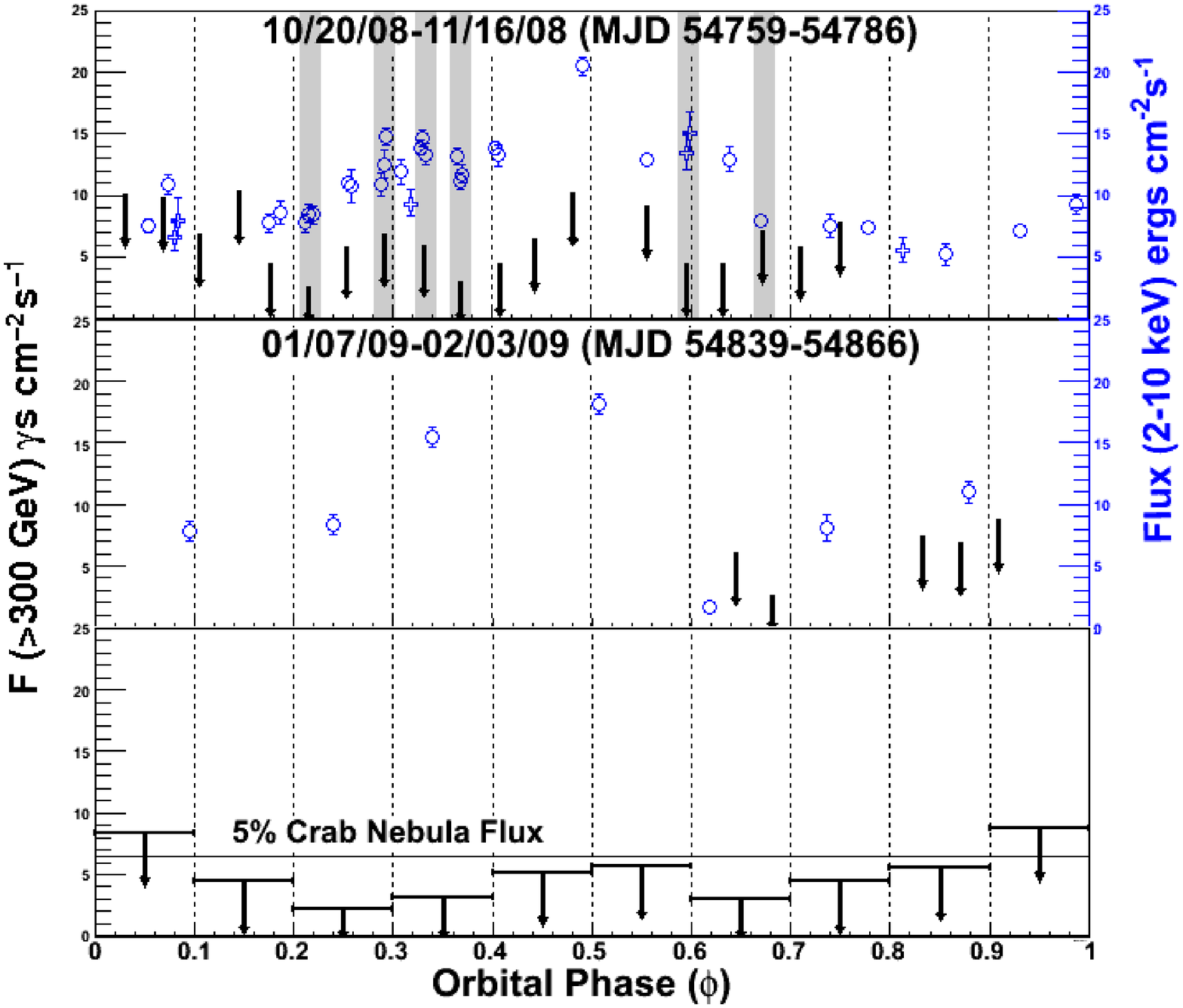}
  \caption{
 Average fluxes or upper flux limits per orbital phase bin for gamma rays with energies above 300 GeV (black points and arrows)
 and X-rays (blue crosses: Swift; blue open circles: RXTE-PCA) 
 from the direction of LS I +61 303 as a function of orbital phase observed from October 2008 to February 2009.  
The bottom panel shows the results averaged over the whole data set; the top panels show the results for individual orbits. Upper flux limits (99\% probability after \cite{Helene-1983}) are shown for data points with significances less than $3\sigma$
 (significance calculation after eq. 17 from \cite{Li-1983}). 
 The shaded regions indicate observations which had directly overlapping X-ray measurements by either RXTE or Swift.
 }
   \label{fig1}
 \end{figure}

\begin{figure}[!t]
  \vspace{5mm}
  \centering
  \includegraphics[width=\linewidth]{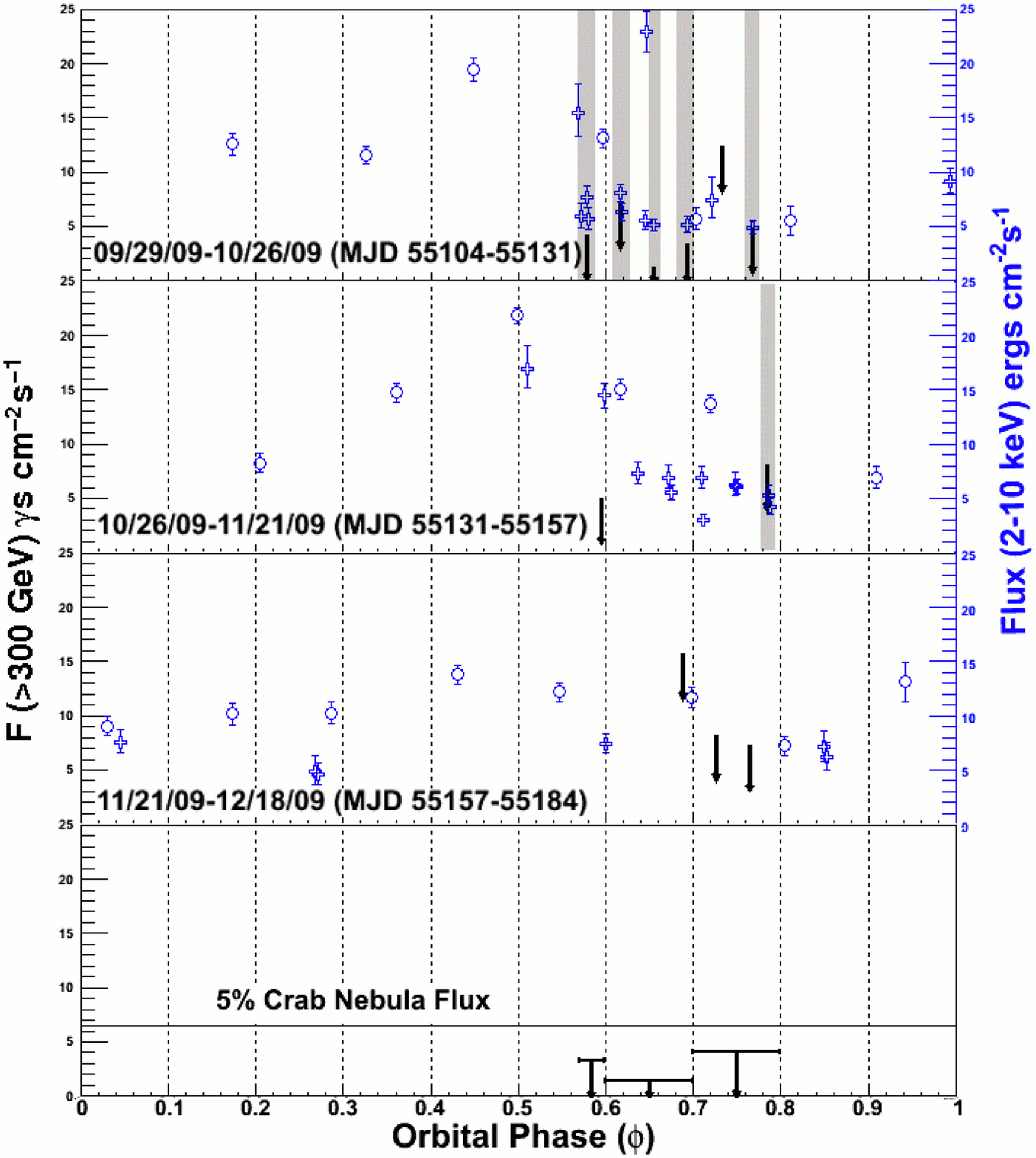}
  \caption{The same as Figure \ref{fig1}, but for the 2009-2010 observing season.
}
   \label{fig2}
 \end{figure}
 
 \begin{figure}[!t]
  \vspace{5mm}
  \centering
  \includegraphics[width=0.9\linewidth]{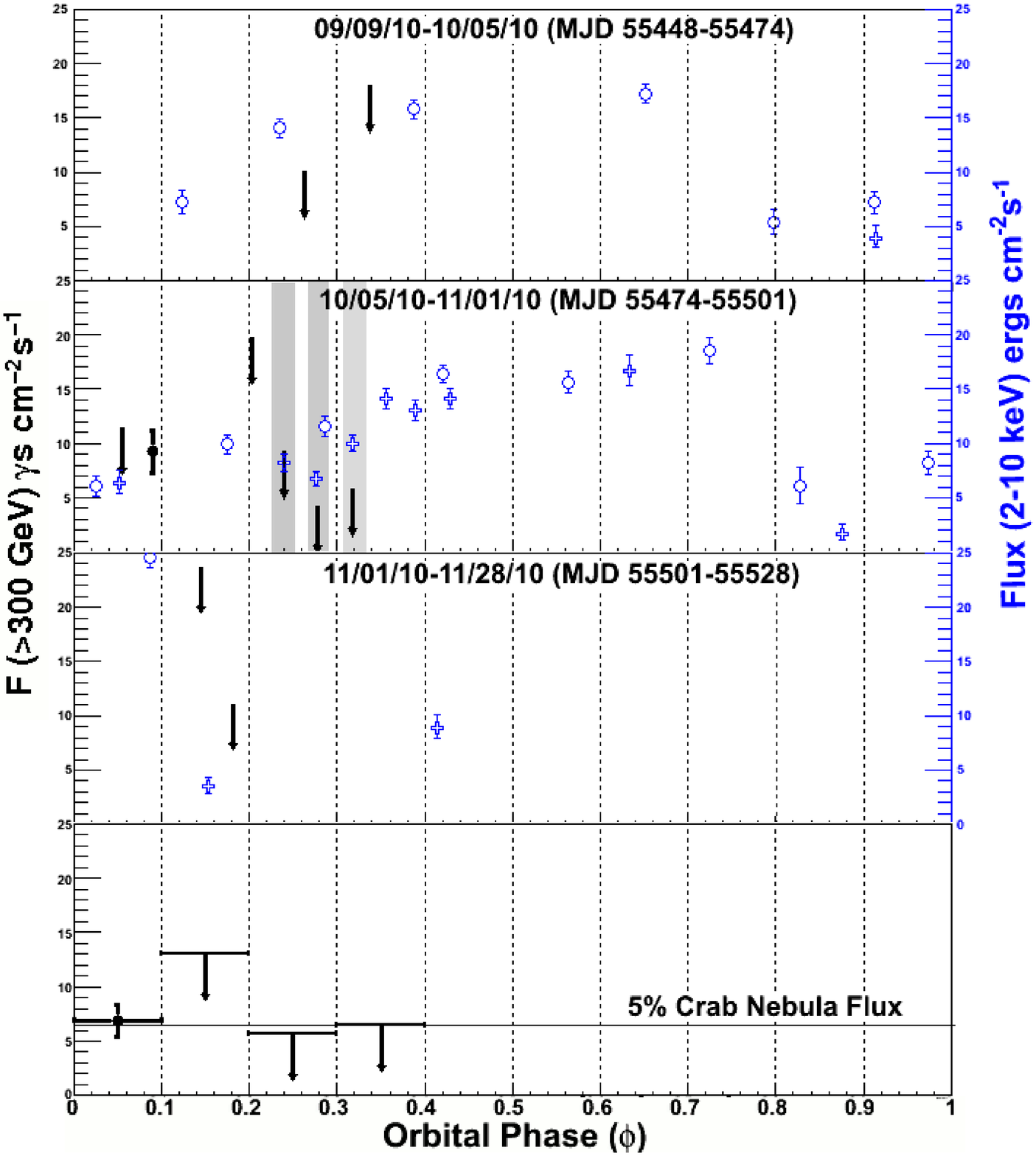}
  \caption{The same as Figure \ref{fig1}, but for the 2010-2011 observing season.
}
   \label{fig3}
 \end{figure}

The high-mass X-ray binary LS I +61 303 consist of a BO-Ve star surrounded by a dense circumstellar disk 
and a compact object of unknown nature (neutron star or black hole). 
Optical observations show that the compact object orbits the star every 26.5 days \cite{Gregory-2002, Aragona-2009}
 on a close orbit, characterized by a semi-major axis of a few stellar radii only.
Periodic variability correlated with the orbital cycle has been observed across the electromagnetic spectrum. 
Long-term X-ray observations show that there is additionally variability of the orbital profiles on multiple time scales,
from individual orbits up to years \cite{Torres-2010}.
Beside the periodic variability, intense X-ray flares with flux increases by several factors on the time scale of seconds
have been observed \cite{Smith-2009}.

Fermi-LAT observations show that the gamma-ray emission at energies about 30 MeV is modulated with the orbital period
(peaking after periastron, $\Phi=0.225$)  \cite{Abdo-2009}.
The energy spectrum shows a sharp exponential cutoff at 6 GeV.
The LAT results reveal overall flux variations not connected to the orbital movement by up to 40\%.
In the VHE range, LS I +61 303 has been detected by ground-based gamma-ray observatories with emission peaking
around apastron \cite{Albert-2006, Acciari-2008}. 
The measured energy spectra for the apastron passage can be described by a power law, 
with a spectral index of 2.4 and a flux above 300 GeV corresponding typically to about 10-15\% of the flux from
the Crab Nebula.

Figures \ref{fig1}-\ref{fig3} show fluxes and flux-upper limits per orbital phase bin
 calculated from 64.5 hours of VERITAS observations between October 2008 and January 2011.
The covered phases by VERITAS observation are influenced by the moon cycle which 
does not allow a complete coverage of a single orbital phase\footnote{Note
that observations taken under moderate moonlight conditions are included in the presented data set.}.
The strategy for the observing season 2008/2009  was to achieve the widest possible coverage
during an orbital phase with a flux sensitivity in the range of 5\% of the flux of the Crab Nebula.
For 2009/2010, the goal was to achieve a deep exposure around apastron in coordination with 
Swift XRT observations.
In the observing season 2010/2011, the phases between superior conjunction and periastron were in the focus
of the observations. 

These results are remarkable in the following ways:
\begin{itemize}
\item No strong evidence for emission from LS I +61 303  during the apastron phases was observed during the observations taken in 2008-2010.
The 2009/2010 measurements, with a deep exposure of about 17 hours in the phase interval $\Phi$=0.5-0.8, result in
upper flux limits (99\% confidence level) of less than 5\% of the flux of the Crab Nebula. 
Note that the object has been detected before at flux levels of  10-15\% of the flux from
the Crab Nebula during these orbital phases.
\item LS I +61 303 has been detected in September-November 2010 
close to the phase of superior conjunction (much closer to periastron passage),
a region of the binary orbit previously undetected by VHE instruments. 
A total of 13.9 hours of LS I +61 303 were accumulated resulting in a post-trials significance of
 5.6 standard deviation.
 The source flux above 300 GeV in the  $\Phi$=0-0.1 phase bin was measured to about 5\% of the flux of
 the Crab Nebula.
 \item The detection of the binary close to superior conjunction does not directly implicate that the processes responsible for
  gamma-ray emission above 300 GeV are different than those at lower gamma-ray energies. 
  The energy spectrum as published by the Fermi LAT collaboration with a cutoff at 6 GeV  is an average spectrum
  for many orbital periods, the data presented here are for a single phase bin at one orbital period.
\end{itemize} 

The light curves from VERITAS show that beside the variability correlated with the orbital cycle, there is
additional orbit-to-orbit variability with no obvious pattern.
The number of upper flux limits and gaps in the light curves shown in Figures   \ref{fig1}-\ref{fig3} 
indicate the two main difficulties for observing LS I +61 303 with current instruments despite the long exposure of more than 60 hours: 
firstly, the orbital period is close to the lunar periodicity (IACTs cannot observe during periods of 
bright moonlight)
and secondly the fluxes away from the apastron regions seem to be close to the sensitivity limits of the
instruments. 
Only continuing monitoring in the X-ray to VHE energy ranges together with well coordinated multiwavelength campaigns
will allow us in the coming years to probe short and long term variability and the emission mechanisms more precisely.

\section{1A0535+262}

 \begin{figure}[!t]
  \vspace{5mm}
  \centering
  \includegraphics[width=0.9\linewidth]{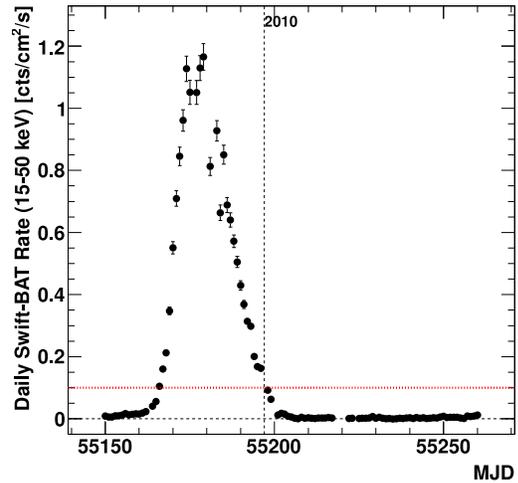}
  \caption{ Swift/BAT counting rate vs time in the 15-50 keV energy range. The horizontal line at 0.1 cts/cm$^2$/s indicates the trigger threshold for observations with VERITAS.
}
   \label{fig4}
 \end{figure}

 \begin{figure}[!t]
  \vspace{5mm}
  \centering
  \includegraphics[width=0.9\linewidth]{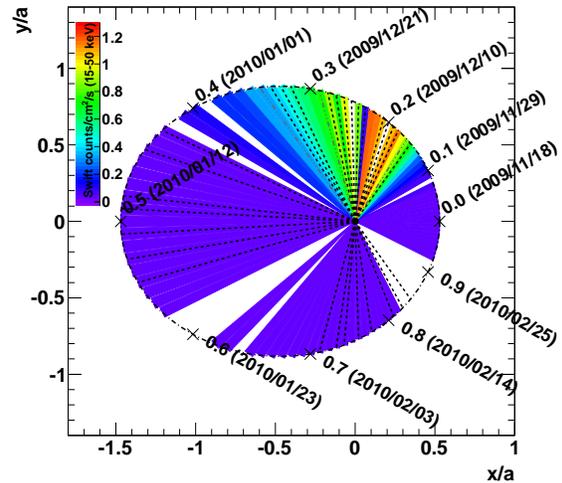}
  \caption{Relative orbit of the neutron star around the Be star in 1A0535+262. The primary star lies in the focus of the ellipse (0,0) and the axis units are multiples of the semi-major axis of the orbit. Note that the inclination of the system is unknown. Indicated in colors is the Swift/BAT counting rate in the 15-50 keV energy range for the orbit starting in November 2009. The dashed lines indicate nights with VERITAS observations, covering the flare, apastron, and periods close to periastron. Orbital parameters after \cite{Coe-2006}.
}
   \label{fig5}
 \end{figure}

1A0535+26 is a nearby (distance of $\approx 2$ kpc) Be/X-ray binary consisting of a pulsar of period 103 s in an eccentric orbit (e = 0.47) with an O9.7-B0 IIIe star.
The orbital period is 111 days.
Giant outbursts at X-ray wavelengths are observed approximately every five years from the direction of 1A0535+262.
They are thought to be associated with the formation of a transient accretion disk \cite{Stella-1986}.

VHE gamma rays are thought to be produced through hadronic or leptonic scenarios.
In hadronic models, protons are accelerated in the magnetosphere of the neutron star and
impact the transient accretion disk.
The pp-interactions in the accretion disk produces VHE gamma rays via the decay of neutral pions \cite{Cheng-1989}.
In leptonic models, electrons are acceleration in strong shocks formed by the interaction  between pulsar wind and stellar wind.
These relativistic electrons may  produce VHE gamma rays by upscattering ambient photons.
The high-energy photons are expected to interact with ambient photons to produce e$^{+}$e$^{-}$ pairs, 
leading to orbital modulation of the VHE gamma-ray flux \cite{Bednarek-1993}. 
The system is therefore opaque to gamma rays near the peak of the outburst, the highest gamma-ray flux is expected
at the beginning or end of the outburst, when thermal emission from the accretion disk is at a minimum \cite{Romero-2001}.

VERITAS observed 1A0535+262 shortly after the onset of the giant outburst in December 2009 for a complete orbital period. 
The Swift BAT measurement in Fig \ref{fig4} shows that the hard X-ray flux reached a level of $>$5 times that of the Crab Nebula. 
The triggering condition for observation of flaring X-ray binaries with VERITAS was fulfilled on 2009 December 5,
observations started on 2009 December 6, shortly after the beginning of the giant flare. They were delayed by one day due to very bright moonlight conditions. The observation covered most of the flare, included the apastron phase, and continued for almost 90 days until the following periastron phase (see Figure \ref{fig5}). 

No evidence for VHE gamma rays has been found for the complete VERITAS data set consisting of 23 hours of high-quality observations.
The flux upper limit at the 99\% confidence level \cite{Helene-1983},
 assuming a power-law-like source spectrum with a spectral index of $\Gamma =  2.5$, is F ($>$ 0.3 TeV) $< 0.5 \times 10^{-12}$
  ph cm$^{-2}$ s$^{-1}$ (about 0.4\% of the flux of the Crab Nebula above 0.3 TeV).

The VERITAS data were arranged in different periods, as gamma-ray production and 
absorption is expected to vary with orbital movement and flaring state. 
The four periods are: rising portion and falling portion of the giant flare, apastron and periastron. 
No VHE emission has been detected in any of these periods; upper limits between 
0.9 and 2.0\% of the flux of the Crab Nebula ($>$ 0.3 TeV) have been derived.

X-ray observations with the Swift XRT and RXTE PCA instruments show that the spectra
are best fitted with a model that consists of blackbody and Comptonized emission from thermal
electrons at temperatures of approximately 2 keV and 6 keV, respectively \cite{Acciari-2011a}.

The non-detection at VHE wavelengths, the results from the spectral analysis of the X-ray observations, and
the non-detection of 1A0535+262 at radio wavelengths indicates that there is no significant
non-thermal electron population in this system. 
The VERITAS upper limit are placing additionally constrains on hadronic models, with the upper limits
 below the predicted flux levels by e.g. \cite{Orellana-2007}.

1A0535+262 represents the archetype of the class of Be binary systems which exhibit giant outbursts 
and the VERITAS observations were the best we could hope to get in terms of coverage and exposure.
A detection or significantly better flux upper limits will probably have to wait for a next-generation
ground-based gamma-ray observatory. 

In summary, extensive observations of the two binaries LS I +61 303 and 1A0535+262 with VERITAS showed
that VHE emission processes in binaries are not well understood. A significant improvement in 
coverage of the different orbital periods and in sensitivity are probably needed to
constrain or refute the large variety of theoretical models.

\subsubsection*{Acknowledgement}

This research is supported by grants from the US Department of Energy, the US National Science Foundation, and the
Smithsonian Institution, by NSERC in Canada, by Science Foundation Ireland, and by STFC in the UK. We acknowledge the
excellent work of the technical support staff at the FLWO and the collaborating institutions in the construction and
operation of the instrument.
G.M. acknowledges support through the Young
 Investigators Program of the Helmholtz Association.

\clearpage

\end{document}